\documentclass[aps,superscript address,preprint,showpacs]{revtex4}
\def\dblone{\hbox{$1\hskip -1.2pt\vrule depth 0pt height 1.6ex width 0.7pt
                  \vrule depth 0pt height 0.3pt width 0.12em$}}
\usepackage{bm}
\usepackage{dcolumn}
\usepackage{epsfig}

\begin{document} 
\title{Green's Function Formalism of Holography with Arbitrary Mass, Spin, and Dimensionality}
\author{Zolt$\acute{{\rm a}}$n Batiz\footnote{zoltan@cftp.ist.utl.pt, visitor}}
\affiliation{\it Centre for Theoretical Particle Physics, 
Instituto Superior Tecnico, 1049-001 Lisbon, Portugal \\ and \\
Companhia dos Mestres,
Rua Ferreira Chaves, no.8 
1070-127 Lisbon, Portugal } 
\author{Bhag C. Chauhan\footnote{chauhan@iucaa.ernet.in}}
\affiliation{\it Dept. of Physics, Govt.. College Karsog, 
Mandi (H.P.) India, 171304}

\date{\today}

\begin{abstract}
In this work we present a mathematical description of 
how one can produce and read a thin hologram. We use different kinds
of waves, such as scalar, vector (electromagnetic field, Maxwell-Proca fields, acoustic waves, etc.). For reading of the hologram, we use the Green's function formalism.
With the help of computer simulations, we investigate the aberrations created by this procedure for the simplest case: 2d-scalar wave case. 
\end{abstract}



\maketitle

\newpage
\section{Introduction}
\label{sec:introduction}
The theory of holography was first developed by Hungarian scientist 
Dennis G$\acute{{\rm a}}$bor around 1947-48 while working to improve the resolution of an electron microscope \cite{gabor}. 
He coined the words hologram and holography from the Greek words holos (whole, entire) and gramma (anything written or drawn). A hologram is defined as the whole [or entire] 
message: the total information. However, holography refers to the information storage process. According to the principle of holography, a detailed three dimensional image of an object can be recorded in a two dimensional photographic film and the image can be reproduced in a three dimensional space. 

In the first holographic experiment Gabor used incandescent light and the results 
were good enough to prove his theory. The quality of the hologram was poor due to the random phase relationships (the noise) produced by the incandescent light. The conventional optical holography is a unique interference pattern of two light beams: A reference beam and an object beam (also known as the diffraction beam) [see Fig. 1]. A laser beam is split by a beam splitter into two parts. The first one (the reference beam) of the divided coherent beam is focused directly on the film and the second one (the object beam) is first flashed onto the object of interest and the modified light waves, after reflection from the object, are then directed on the film where they interact with the reference beam. The interaction of the coherent information in the reference beam and the object beam creates the interference pattern and is recorded (encoded) in the film emulsion. The complex patterned information stored in the film is called 'hologram'. When the developed film is again illuminated by a coherent light beam, the encoded information is projected into local space and an image of the original object is reconstructed.
\begin{figure}[hbt]
\centerline{\epsfig{file=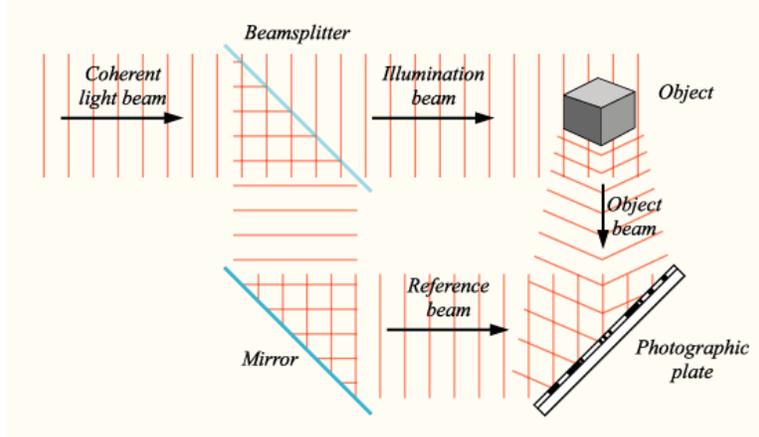,width=4.0in}}
\caption{Image taken from \cite{wiki}}
\label{fighbt}
\end{figure}

The hologram can be produced by any wave kind of wave action \cite{BW}. The conventional laser holography \cite{conven}, acoustical holography \cite{acoust} and electron holography \cite{electron} came up about fifty years after the first articulated hologram theory. 
Due to an advancement in the computer technology the computational holography \cite{comp}, a computer synthesized real-time interactive or virtual reality display of holograms, is a growing area of interest. 

In this work we investigate the aberrations caused by the holographic procedure. We also studied the effect of reading beam hitting the screen not in the same direction as the reference wave, but in a slightly different one, and by using slightly different wavelengths for the reading, etc. As it is known in optics \cite{BW}, the image of a point is rarely a point, but a fuzzy object. We investigate the width of this image and its dependence on the size of the holographic screen, the distance between the screen and the object, the distance of the object from the symmetry axis of the screen, the nature of the mapping and reading wave. By 'waves of different nature', we mean scalar, spinor, vector and tensor waves etc... 

In the section-\ref{S:form} we describe the theory of creation and reading of a hologram using various types of waves in the usual space and hyperspace. In the section-\ref{S:stationary point} we use computer simulations and study the image created by the holographic mapping of a single stationary point with scalar wave in 2d-space, and 
then reading the hologram. In this section we also investigate the aberrations caused by the holographic procedure. In the section-\ref{S:conclusions} we present the discussion and conclusions.
 
\section{Holographic Mapping and Reading}
\label{S:form}
In this section we first describe the creation of holograms when the object is in the usual three-dimensional space (subsection \ref{S2:form}). In the present study the hologram is two-dimensional. This of course can be generalized also for a different dimensionality 'd', but then the dimensionality of the hologram will be 'd-1'.

Next, we describe the reconstruction of the holographic image in the d-dimensional space (subsection \ref{S1:form}). As said above, the holograms can be formed in the presence of any wave action \cite{BW}, first we use scalar fields and we generalize our procedure for several other fields, such as vector (electromagnetic, acoustic, Maxwell-Proca), tensor (gravitational), and spinor fields. 
We also describe how can one generalize these calculations to a space of different dimensionality.

To generalize the results of our previous work \cite{bc}, here we use a massive scalar field and we prove that for this the Green's functions found in the classical text books of electricity and magnetism, such as \cite{eyges} are 
adequate. Scalar waves can be produced by using scalar particles, or pseudo scalars, as in the case of our calculation space reflection is not used. Even if we make use of electromagnetic waves, that are vectors, they can be regarded as scalars in the so-called paraxial approximation, meaning that the size of the screen is much smaller than the object-screen distance \cite{lb}. But in order to create and read holograms, one must  investigate how different kind of waves are reflected and how do they affect photosensitive materials. 
We restrict our discussion to reflection holograms, although the generalization to transmission holograms is straightforward, and we assume thin holograms only.

\subsection{The Physics of Different Kinds of Waves}
\label{S0:form}

For scalar waves, there is obviously no issue of polarity. For vectorial waves with several components $A^i$, the reflection process can be modelled as
\begin{equation}
A^{i \prime}=-R T^{ij}A^j,
\label{eq1}
\end{equation}

where the prime referring to the components after reflection, R is the reflectivity, and $T^{ij}$ is the reflection tensor. The negative sign is due to the phase the phase shift of  $\pi$ due to reflection. For tensor waves, this is generalized as
\begin{equation}
A^{i j \prime}=-R T^{ik}T^{jl}A^{kl}.
\label{2}
\end{equation}

Next we describe how this transformation tensor is determined in the case when we use vectorial waves. The wave vectors of the incident and reflected waves are $\vec{k}_i$ and $\vec{k}_{refl}$. When we create the hologram, the first wave vector refers to the object beam before it hits the object, while the second the wave vector of the same when it leaves the object. However, when we read the hologram, the first 
wave vector is that of the reference beam and the second is that of the wave 
reflected from the hologram.  We assume that they are equal in magnitude, 
$|\vec{k}_i|=|\vec{k}_{refl}|=k$, i.e. one has elastic scattering.
These two vectors determine a plane of incidence. We then construct the unit vectors ${\hat v}_i$ 
and ${\hat v}_r$, requiring that both be in the plane of incidence 
and $\vec{k}_i \cdot {\hat v}_i=\vec{k}_{refl} \cdot {\hat v}_r=0$, therefore they are found to be
\begin{equation}
{\hat v_i}=\frac{-\vec{k}_i \cos{\theta}+\vec{k}_{refl}}{k \sqrt{1-\cos^2{\theta}}}
\end{equation}

and
\begin{equation}
{\hat v_r}=\frac{\vec{k}_i -\cos{\theta} \vec{k}_{refl}}{k \sqrt{1-\cos^2{\theta}}},
\end{equation}

where $\cos{\theta}=\vec{k}_i \cdot \vec{k}_{refl}/k^2$ is the cosine of the angle between the two wave vectors.

In the electromagnetic case (whether the wave is transverse as in the case 
of free waves, or longitudinal or mixed, as they can appear in plasmas) as 
well as in the case of massive vector bosons, that are called Maxwell-Proca waves, 
the reflection from perfect conductors has the same boundary conditions. 
We model our reflection tensor in Eq. (\ref{eq1}) to satisfy these boundary conditions, which are compatible with the very reasonable and commonly used assumption that there will be a phase shift of $\pi$ due to reflection.
This mean that the polarization component that is perpendicular to the plane of incidence will change sign (which is signified by the 'minus' sign in Eq. (\ref{eq1}), but we later include this sign in the phase).
The polarization component that is parallel to $\vec{k}_i$ will be parallel to $\vec{k}_{refl}$ and will also change sign, just like the component parallel to $\hat v_i$ that will be parallel to $\hat v_r$. These considerations will give the following form to the tensor $T$
\begin{equation}
T^{ml}=\delta_{ml}-{\hat k}_i^m.{\hat k}_i^l+{\hat k}_{refl}^m.{\hat k}_i^l
-\hat v_i^m .\hat v_i^l+ \hat v_r^m.\hat v_i^l .
\label{3}
\end{equation}

The ${\hat k}_i$ and ${\hat k}_{refl}$ symbols mean the unit vectors 
belonging to ${\vec k}_i$ and ${\vec k}_{refl}$.

One can verify that a reflection of this kind preserves helicity.
While modelling spin we consider processes that preserve helicity in order to be consistent with our former results and we also assume that the reflection tensor does not modify the norm, since we want to incorporate 
this effect into $R$, and we also assume that there is a phase shift of $\pi$.

Therefore Eq. (\ref{eq1}) is maintained, while the transformation tensor is different
\begin{equation}
T^{ml}=a_0 \dblone+i \vec b . \vec \sigma_p,
\label{4}
\end{equation}

with

\begin{eqnarray}
a_0=&&\frac{1}{2} \sqrt{1+\cos{\theta}} , \\ \nonumber
b_j=&&\frac{1}{2} \sqrt{1+2{\hat k}_i^j . {\hat k}^j_{refl}-\cos{\theta}},
\label{5}
\end{eqnarray}

and the $\sigma_p$ symbols stand for the usual Pauli matrices.

The only link missing from our discussion is the investigation of the manner these waves interact with photographic materials.
It is known that for optical holograms the maxima correspond to the ventral points 
of the electric fields and not those of the magnetic fields and that the electric fields ${\vec E}$ are parallel 
to the vector potential ${\vec A}$: ${\vec E}=i \omega {\vec A}$ \cite{lb}.
Therefore we model all our pure transverse waves in the same way as the free electromagnetic waves 
and the purely longitudinal ones as electrostatic waves. For these waves, if the intensity $I$ is defined as
$I={\vec E}^2$ one can see that $I  \sim {\vec A}^2$. 

However, maintaining the same definition of $I$ for the Maxwell-Proca case, one does not have the same proportionality between intensity and the square of the vector potential as  in the electromagnetic case. It is known \cite{jr} that for massive vector bosons the gauge symmetry is broken and the Lorentz condition is mandatory, therefore 
besides the transverse fields ${\vec A}_{\perp}$ there will be longitudinal components 
${\vec A}_l$ even in the case of free waves. The frequency $\omega$ will not only depend on the 
wave vector, but also on the mass of the particle $M$: $\omega=\sqrt{M^2+k^2}$, in the so-called natural system ($\hbar=c=1$). For these kind of waves it is also crucial to find a connection between the ${\vec E}$ and ${\vec A}$ fields, since the former is involved in the computation of the intensity, while the latter is given by the Green's function formalisms directly, as we see in subsection \ref{S1:form}.

Maintaining the same definition for the electric fields as in Electrodynamics and making use of the Lorentz condition, we find that the electric field is ${\vec E}=ik{\vec A}_{\perp}+\frac{iM^2}{\omega}{\vec A}_l$, 
therefore the proportionality relation becomes 
\begin{equation}
I \sim \left| {\vec A}_{\perp}+\frac{M^2}{\omega k}{\vec A}_l \right|^2.
\label{5a}
\end{equation}

In the scalar case we assume that the intensity is proportional to the absolute value squared of the wave field and in the spinor and tensor case this will be proportional to the sum of the absolute values squared of each component of the wave field.

\subsection{Making the Hologram}
\label{S2:form}

The hologram is produced, as seen in Fig.(1), by splitting a single beam into 
two pieces: the object beam with its wave vector $\vec{k_o}$ and the
reference beam directly shed on the photographic plaque with a wave vector $\vec{k_r}$.
As we have said before, we assume a d-dimensional space with the holographic screen 
lying on the 'd-1' plane by construction. 
 
We assume that the phase of the reference beam is $\phi_{r_0}$ while the phase of object beam is $\phi_{o_0}$ in the origin, which is also the geometrical center of the holographic screen.  
The wavelength of the radiation is $\lambda=2 \pi/|{\vec k}|$, where ${\vec k}$ is the wave vector.
Therefore the phase of the reference beam at an arbitrary point $A$ of the screen (whose position vector is ${\vec r}_A$) is $\phi_r(A)=\phi_{r_0}+{\vec k} \cdot {\vec r}_A$. Likewise if the position vector of the object is ${\vec r}_o$, the phase of the object beam when it hits the object is $\phi_o=\phi_{o_0}+{\vec k} \cdot {\vec r}_o$. When the object beam is reflected, 
it acquires an additional phase $\pi$, and when it hits the screen at $A$, its phase will be  
$\phi_o(A)=\phi_{o_0}+{\vec k} \cdot {\vec r}_o+\pi+k|{\vec r}_A-{\vec r}_o|$.
In the case of reflection, we added a phase $\pi$, which is 
not always realistic, but in most cases is a good approximation and is widely used in eikonal optics \cite{BW}.

The wave field that is reflected from the object and then hits the screen has a value of $E_{o}(A)$, which is 
\begin{equation}
{E}_{o}(A)=E_{o0}\sqrt{\sigma R} \frac{\exp{(i\phi_o(A)})}{\sqrt{4 \pi} |{\vec r}_A-{\vec r}_o|}.
\label{1aeq1a}
\end{equation}

We labeled the amplitude of the object wave by $E_{o0}$, the cross section of the object by $\sigma$, its reflectivity by $R$, and we assumed isotropic reflections, which is also a reasonable approximation.
Likewise the reference beam (whose amplitude is $E_{r0}$) will have its contribution to the field on the screen
\begin{equation}
{E}_r(A)=E_{r0} \exp{(i \phi_r(A))}. 
\label{1aeq1}
\end{equation}

The fields ${E}_{o}(A)$ and ${E}_{r}(A)$ can be added together and squared, and then we have the interference picture that is the 'hologram'. How would one generalize for an arbitrary number of dimensions?
Only Eq. (\ref{1aeq1a}) will be modified, while the phases will be given by the same formulae as before, and Eq. (\ref{1aeq1}) is also maintained, with the exception that the dot product will contain a different number of terms.

As we have found previously \cite{bc}, in the case where we assume isotropic reflections and the number of dimensions 'd' is greater than two, the reflected wave field from the object will read
\begin{eqnarray}
{E}_{o}(A)=E_{o0}\sqrt{\sigma_d R \frac{\Gamma(\frac{d}{2})}{2 \pi^{d/2}}}
\frac{\exp{(i\phi_o(A))}}{|\vec{r_A}-\vec{r_o}|^{d/2-1}} ,
\end{eqnarray}

where $\Gamma$ is the Euler function and we also have to use the generalized cross section $\sigma_d$  instead of $\sigma$. In this case, as we have said previously, the dimensionality of the holographic picture is 'd-1'.

The only infra-dimensional case which makes sense is 'd=2', where the hologram is one-dimensional and Eq. (\ref{1aeq1a}) will be modified as
\begin{eqnarray}
{E}_{o}(A)=E_{o0}\sqrt{\sigma_d R} 
\frac{\exp{(i\phi_o(A))}}{\sqrt {2 \pi|\vec{r_A}-\vec{r_o}|}} .
\end{eqnarray}

This discussion covers the effects of the dimensionality while creating the hologram. 
As we have seen, the velocity does not show up in these formulae, the only thing that 
is referring to the wave is the wave vector, therefore there is no dependence on the mass except for the case of the Maxwell-Proca waves due to Eq. (\ref{5a}). This will be true also in the case of hologram reading.

We can generalize the former discussion for the case of waves with different nature, which implies a different number of components. For vector and spinor fields, there will be $i$ components $E^i(A)={E}^i_{o}(A)+{E}^i_{r}(A)$ where 
\begin{equation}
{E}^i_{o}(A)=E^{i \prime}_{o0} \frac{\exp{(i\phi_o(A))}}{\sqrt{4 \pi} |{\vec r}_A-{\vec r}_o|}
\label{1aeq1ai}
\end{equation}

and
\begin{equation}
{E}^i_r(A)=E^i_{r0} \exp{(i \phi_r(A))}, 
\label{1aeq1i}
\end{equation}

where $E^{i \prime}_{o0}$ are the amplitudes of the different components for the 
object beam when it leaves the object, while $E^i_{r0}$ are the 
different components for the reference beam. The intensity on the screen at point $A$ will be $I(A)=\sum_i{|{E}^i(A)|^2}$. The components $E^{i \prime}_{o0}$ will be evaluated as follows
\begin{equation}
E^{i \prime}_{o0}=-\sqrt{\sigma_d R } \sum_j T^{ij} E^j_{o0}.
\label{1spinor}
\end{equation}

\subsection{Hologram Reading} 
\label{S1:form}
The hologram is read once one sheds a wave on it. It may be noted that the direction and the 
frequency of this beam should be the same as those of the reference beam 
that was used to create it. 

In order to determine the image generated by a hologram when we 
shed some wave onto it, one must know the reflected fields at any given point. 
The intensity of these fields is related to the squares of the  
reflected wave fields as we have described in subsection \ref{S0:form}, 
and knowing that, we can have an analytical description of the generated image.
If some wave is reflected from a surface (such as a hologram), we can
compute the reflected fields on the surface and we can examine how those fields propagate. First we consider how static fields are determined from known boundary conditions and after that we extend the calculation for wave fields. For simplicity, we start with the (massive) scalar case.

The Green's function $G({\vec r})$ of a static field is defined as
\begin{equation}
\left( {\nabla}^2-M^2 \right) G({\vec r})=-\delta({\vec r}).
\label{1eq1}
\end{equation}
The field at any given point can be calculated as 
\begin{eqnarray}
\varphi({\vec r}^{\prime})=\int_V d^3 {\vec r} 
\varphi({\vec r})\delta({\vec r},{\vec r}^{\prime})=
-\int_V d^3 {\vec r} \varphi({\vec r}) 
\left({\nabla}^2 -M^2 \right) G({\vec r},{\vec r}^{\prime}),
\label{1eq2}
\end{eqnarray}

where the derivative ${\nabla}$ is related to the variable ${\vec r}$.

After applying the following identities
\begin{eqnarray}
\varphi({\vec r}) 
{\nabla}^2 G({\vec r},{\vec r}^{\prime})=&&{\vec {\nabla}}
\left[ \varphi({\vec r}) {\vec {\nabla}}G({\vec r},{\vec r}^{\prime}) \right]
-({\vec {\nabla}}G({\vec r},{\vec r}^{\prime}))
({\vec {\nabla}}\varphi({\vec r})) \, ,\nonumber \\ 
({\vec {\nabla}}G({\vec r},{\vec r}^{\prime}))
({\vec {\nabla}}\varphi({\vec r}))=&&
{\vec {\nabla}}(G({\vec r},{\vec r}^{\prime}) 
\nabla \varphi({\vec r}))-G({\vec r},{\vec r}^{\prime}) 
\nabla^2 \varphi({\vec r})\, ,
\label{1eq2a}
\end{eqnarray} 

and making use of the Laplace equation
\begin{eqnarray}
\left( \nabla^2 -M^2 \right) \varphi({\vec r}))=-\rho({\vec r})\, ,
\label{1eq3}
\end{eqnarray}
the field in any point can be calculated as
\begin{eqnarray}
\varphi({\vec r}^{\prime})=\int_V d^3 {\vec r} 
G({\vec r},{\vec r}^{\prime})\rho({\vec r}) + \int_S d^2 {\vec r}
G({\vec r},{\vec r}^{\prime}) {{\nabla}_n}\varphi({\vec r})-
\int_S d^2 {\vec r}
\varphi({\vec r}){{\nabla}_n}G({\vec r},{\vec r}^{\prime}))\, ,
\end{eqnarray}

where ${{\nabla}_n}$ is the component of the derivative that is perpendicular
to the surface.
 
The first term refers to the sources and we assume that there 
are no sources in the part of space we examine. The other two terms are 
the so-called surface terms. In Optics, these are called Kirchhoff integrals. 
Whenever the first surface term vanishes(and the Green's function must be chosen accordingly, so that it vanishes on the surface) we must know the value of the field on the surface, and we are said to use the Dirichlet conditions.

If we know only the derivatives of the fields on the surface,
we must require that the normal derivative of the Green's function vanishes on the surface, and we are said to make use of a Neumann Green's function. Note that any of these conditions can be met at any time (although not both at the same time) because Eq. (\ref{1eq1})
does not completely fix the Green's function, so we might add any term whose Laplacian is zero 
(in the region of space we are interested in) in such a way that the new Green's function satisfies either one of the two conditions. Because we can calculate the fields at the surface, we use a Dirichlet Green's function, so our field at any given point is expressed as follows
\begin{eqnarray}
\varphi({\vec r}^{\prime})=-\int_S d^2 {\vec r}
\varphi({\vec r}){{\nabla}_n}G({\vec r},{\vec r}^{\prime}))\, .
\label{1eq3a}
\end{eqnarray}

From the image solution for the auxiliary electrostatic problem, the Green's 
function for Dirichlet conditions can be calculated.
We need to know the Dirichlet Green's function 
on a plane. First we define some new variables ${\vec r}_{1}$ and 
${\vec r}_{2}$ as 
${\vec r}_{1,2}=(x-x^{\prime}) {\vec e}_1+
(y-y^{\prime}){\vec e}_2+(z \mp z^{\prime}){\vec e}_3$, 
(in terms of our orthonormal basis ${\vec e}_1$, ${\vec e}_2$, ${\vec e}_3$),
$r_{1,2}=\sqrt{({\vec r}_{1,2} \cdot {\vec r}_{1,2})}$.

In these terms, the Dirichlet Green's function is given in \cite{eyges} as  
\begin{equation}
{\tilde G({\vec r},{\vec r}^{\prime})}=\frac{1}{4 \pi}
\left( \frac{1}{r_1}-\frac{1}{r_2} \right)\, .
\label{1eq4}
\end{equation}

If instead of static field we have wave fields, this Green's function is replaced with 
\begin{equation}
G({\vec r},{\vec r}^{\prime};t)=\frac{1}{4 \pi}
\left( \frac{\delta(t-r_1/v_f)}{r_1}-\frac{\delta(t-r_2/v_f)}{r_2} \right)\, ,
\label{1eq5}
\end{equation}

where $v_f=\omega/k$ is the phase velocity of our wave. Since we consider only one frequency ($\omega$), we only need the Fourier transform of this Green's function, which is
\begin{equation}
G({\vec r},{\vec r}^{\prime};\omega)=\frac{1}{4 \pi}
\left( \frac{\exp{(ikr_1)}}{r_1}-\frac{\exp{(ikr_2)}}{r_2} \right)\,.
\label{1eq6}
\end{equation}

Note that the dependence on $\omega$ and hence on the mass has disappeared.
If we change the sign between the two terms of the RHS of the former equation, 
we get the Neumann's Green's function. If we drop the second term, we need to know both the field and its derivative on the integration surface, which is another way we could proceed. 
We choose to use the Dirichlet's Green's function, however. 
Now we substitute Eq. (\ref{1eq6}) into Eq. (\ref{1eq3a}), but how one can justify this substitution, since Eq. (\ref{1eq3a}) has been derived based on the assumption that the fields are static. 
Let's check this in the following way: we know if there is a wave field, the Eq. (\ref{1eq3}) is replaced with
\begin{equation}
\left( {\nabla}^2 -\frac{\partial^2}{\partial t^2}-M^2 \right) 
\varphi({\vec r})=-\rho({\vec r,t}).
\label{1eq7}
\end{equation}

On the other hand, if we consider one frequency and retarded waves only, our field and source can be expressed as 
\begin{eqnarray}
\varphi({\vec r},t)=&&\varphi({\vec r},t=0)\exp{[-i \omega(t-l/v_f)]},
\nonumber \\
\rho({\vec r},t)=&&\rho({\vec r},t=0)\exp{[-i \omega(t-l/v_f)]},
\label{1eq8}
\end{eqnarray}

where $l$ is the distance between the source and observer.
Now, substituting this into Eq. (\ref{1eq7}) and dividing the resulting equation by
$\exp{(-i \omega(t-l/v_f))}$ we obtain Eq. (\ref{1eq1}).
So if we work with the time Fourier transforms of the wave fields and Green's functions and assume only one frequency, we are able to make use of the static formulation of the problem using the Fourier transform of the Green's function we have just given in Eq. (\ref{1eq6}).

The normal derivative of this Green's function on the surface defined by the hologram is
\begin{equation}
{\nabla}_n G({\vec r},{\vec r}^{\prime})=
-\frac{\partial}{\partial z}G({\vec r},{\vec r}^{\prime})\vert_{z=0}.
\label{1eq9}
\end{equation}

If the dimensionality 'd' were different (but $d>3$), our Green's function would be modified as
\begin{eqnarray}
G({\vec r},{\vec r}^{\prime})=
\frac{\Gamma(\frac{d}{2})}{2 \pi^{d/2}} \left[
\frac{\exp{(i k r_1)}}{r_1^{(d-1)/2}}- \frac{\exp{(i k r_2)}}{r_2^{(d-1)/2}} \right],
\label{1eq10a}
\end{eqnarray}

while for $d=2$, (the only meaningful case of an infra space in holography) is
\begin{eqnarray}
G({\vec r},{\vec r}^{\prime})= const. \times \left[\exp{(i k r_1)}
\log{ \left( \frac{r_1}{R_0} \right) }- \exp{(i k r_2)}
\log{ \left( \frac{r_2}{R_0} \right) } 
\right],
\label{1eq10aa}
\end{eqnarray}

where $R_0$ is arbitrarily fixed in order to make the Green's function vanish when $r_1=r_2=R_0$. The constant in the former equation is not even important.

Now the only thing left to be determined is the reflected field at any given point of the hologram.
For scalar waves, the phase and amplitude of the reflected wave depends on the phase and amplitude of the reading wave
\begin{eqnarray}
E^{\prime}_r(A)=-|R_h| E_r(A)\, 
\label{1eq13}
\end{eqnarray}

where $E^{\prime}_r(A)$ is the reflected wave, $R_h$ is the reflectivity of the 
hologram (it is the hologram data file generated in the previous step) 
and $E_r(A)$ is the reading wave (exactly same as the reference beam in 
Eq. (\ref{1aeq1})).

As in the previous section, the field of the wave that hits the screen is
\begin{eqnarray} 
E_r(A)=E_{r0} \exp{ \left[ i \left(\phi_{r_0}+{\vec k} \cdot {\vec r}_A \right) \right]},
\label{1eq14}
\end{eqnarray}

which we introduce into the Eq. (\ref{1eq13}). In the case of wave fields with several components, one must use the appropriate relations for the reflection as given in subsection \ref{S0:form}.

We incorporate Eqs. (\ref{1eq3a}), (\ref{1eq10aa}), (\ref{1eq13}) and (\ref{1eq14}) into a numerical code to compute the reflected fields (and therefore the intensities in the manner presented in subsection \ref{S0:form}) at any given point of the space. Therefore, after reading we obtained the image from the hologram.

\section{Computer Generated Holography}
\label{S:stationary point}

In this section we generate a computer code which consider a stationary point-like 
object of negligible but finite physical size for study. In the simulation we use waves of 
different types, like scalar, spinor, vector and tensor waves, for creating and reading the hologram in various dimensionality.

Like any optical procedure \cite{BW}, the process of generating and creating the hologram too will give us a blurry picture instead of a single point. In this section we are empirically studying the dynamics of this blurriness and try to read some physics into it. 

For the whole process we assume that the point-like object is situated in a 'd' dimensional space and the hologram is created in a 'd-1' dimensional space. The object is separated from the center of the rectangular screen (whose picture is the actual hologram) by a distance 'DD'. The sides of the holographic screen are equal and their size is 'D', and that the line of separation is perpendicular to the screen.

There is no need to play with wavelength because it appears in ratios with image width, distance from the screen, and off-centricity. So for convenience we fixed the wavelength $\lambda=1$ unit and thus wave vector $k=2\pi$. The geometrical size of the point is taken as $S_1=5\times 10^{-5}$ units.

For the numerical integration that is involved in creating and reading the hologram, we divided the screen into several thousands ($N_P^{2}$) of equal regions (pixels); say $10000 \times 10000$. The number of necessary grid-points was estimated as follows: $N_P$=Screen Size*Number of Fringes per unit Wavelength.
We estimated the size of the fringe, knowing this, we computed the number of fringes per screen size, and placed 10 points per fringe, then, in order to check stability we doubled the number of grid-points and so on.
 
We played with the parameters; screen dimensions (D), distance of image centre from the screen (DD), number of grid-points ($N_P$), directions of reference and object beams and off-central configurations etc... to see the effect on best possible image of the object. We also check the effect due to the change in wavelength $\lambda$. We tabled down the width of image $(\Gamma)$ with respect to these parameters.

\begin{figure}[hbt]
\centerline{\epsfig{file=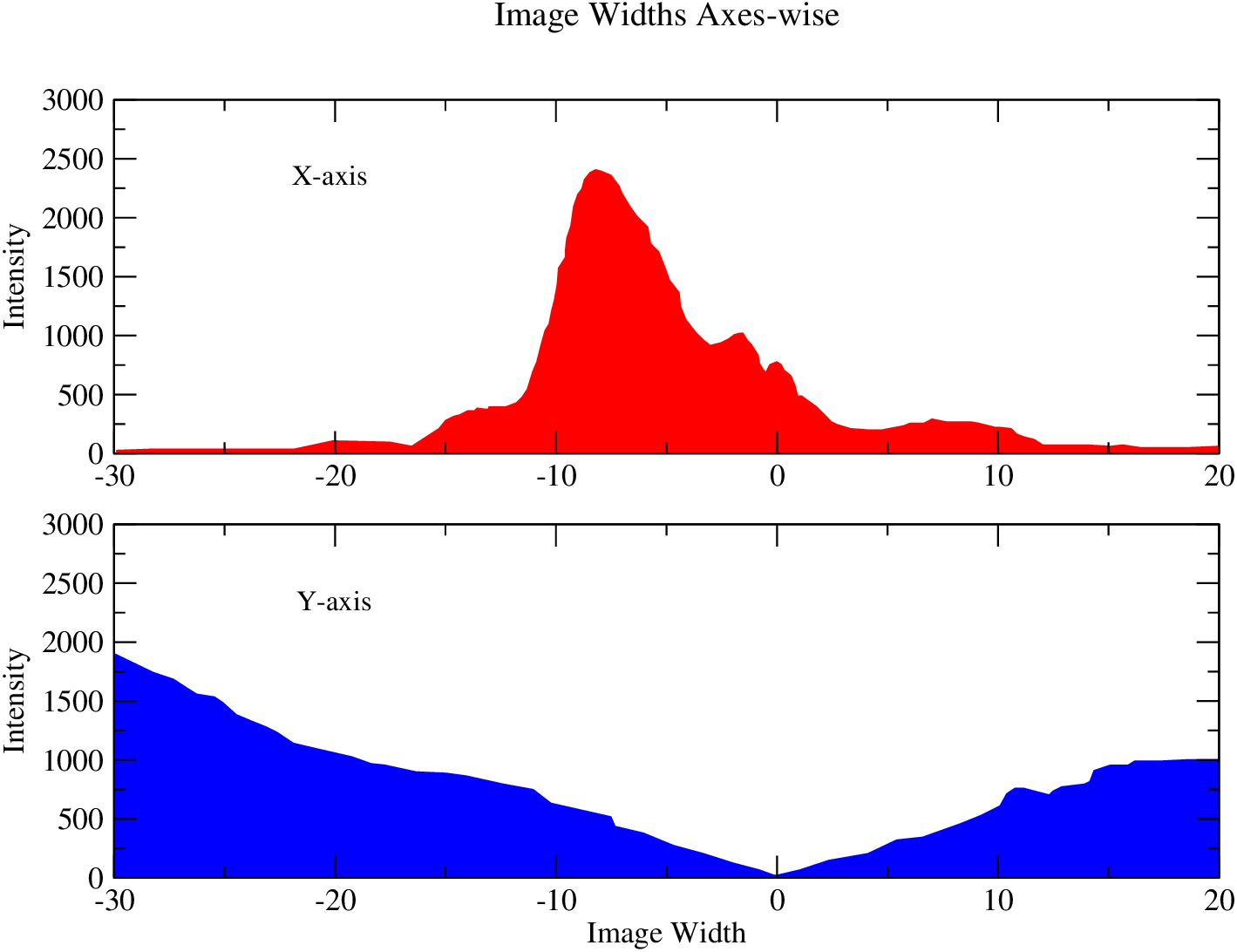,width=4.0in}}
\caption{The image for $DD=500$ units, $D=10$ units
and $\lambda=1.0 $ units.}
\label{fig1}
\end{figure}

In this work we consider the simplest case of scalar wave and a point-like object in a two-dimensional (d=2)-space. The infra-space hologram thus created is one (d=1) dimensional space. The direction cosines of the object and reference beams are taken slightly different; ($co(1)=0.2$) and ($cr(1)=0.22)$, respectively.
The numerically created hologram of the object thus is read back in the two-dimensional space.
We noted the image has a sharp peak, centred at $-8$ units, along x-axis, however it is quite blurred along -ve side of the y-axis, as shown in Fig (\ref{fig1}). No effect in the image width and noise with variation in the distance of object from the screen has been found as shown in the Table I.
\begin{table}
\begin{tabular}{|c|c|c|c|c|} \hline 
$DD$  & 5.0  & 50.0  & 500.0 & 5000.0   \\ \hline
$\Gamma$ & 6.0 & 6.0 & 6.0 & 6.0  \\
  \hline
\end{tabular}  
\caption{\it  Empirical study of the parameters in units: $\Gamma$ v/s $DD$. Here $D=60.0$ units.}
\label{tab1}
\end{table}

In Table II we fixed the centre of image formation $DD$ at 500 units and study the variation of the screen width $D_1=D_2=D$. We found that $\Gamma=6.0$ units. We also noted that in most of the other cases there appear double peaks intermingled with each other. 
 
\begin{table}
\begin{tabular}{|c|c|c|c|c|c|c|c|} \hline 
$D$  & 5.0  & 10  & 12 & 15 & 20 & 30 & 40   \\ \hline
$\Gamma$ & 13 & 6.0 & 12 & 13.5 & 19 & 26 & 37  \\
\hline
\end{tabular}  
\caption{\it  Empirical study of the parameters in units: $\Gamma$ v/s $D$, with $DD=500$ units.}
\label{tab1a}
\end{table}

\vspace*{1.1cm}
We present a case where the screen width is fixed at $D=10$ units and we study the dependence of image width with the variation of the distance from the screen as depicted in Table III.

\begin{table}
\begin{tabular}{|c|c|c|c|c|c|c|c|c|} \hline 
$DD$  & 5  & 10  & 20 & 30 & 40 & 50 & 60 & 70  \\ \hline
$\Gamma$ & 5 & 6 & 8 & 10.5 & 12 & 15 & 17 &  18.5 \\
  \hline
\end{tabular}  
\caption{\it  Empirical study of the parameters in units: $\Gamma$ and $DD$, with $D=10$ units.}
\label{tab1b}
\end{table}

\begin{figure}[hbt]
\centerline{\epsfig{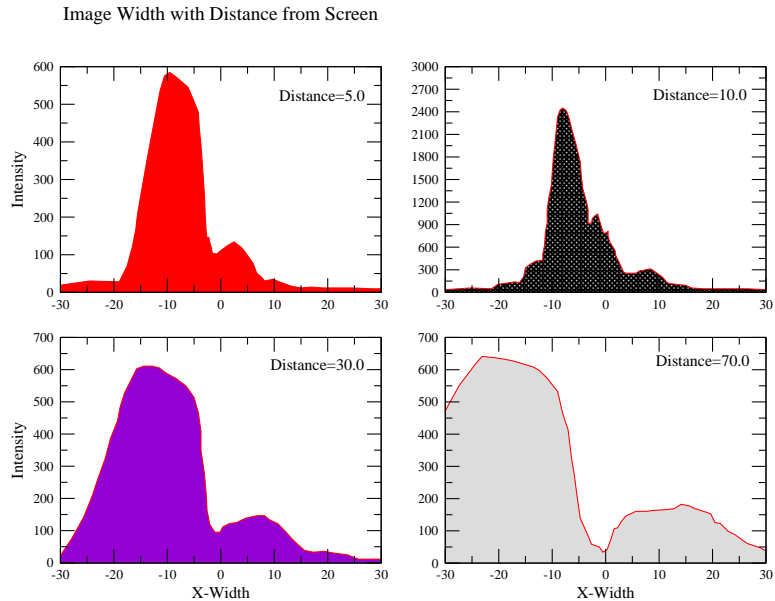}}
\caption{The image for $DD=10$ units, $D=5$ units
and $\lambda=1.0 $ units.}
\label{fig2}
\end{figure}

In the Fig. (\ref{fig2}) we plot image intensity versus x-width of the image in four panels for $DD=5.0,~10.0,~30.0,~70.0$ units, respectively. One can see from the Figure that the blurriness is minimum for $DD=10$ units, in other words the image is focussed at a distance of 10 units.

\vspace*{1.5cm}

\begin{figure}[hbt]
\centerline{\epsfig{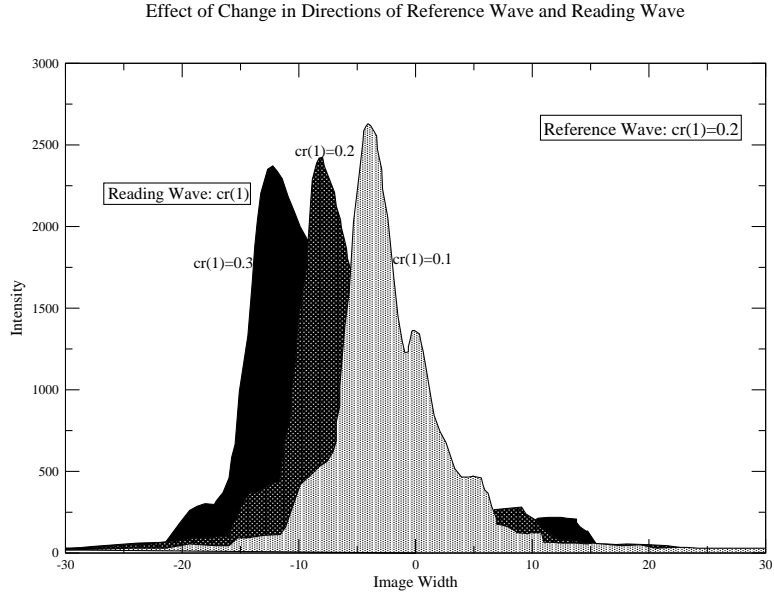}}
\caption{The image for $DD=10$ units, $D=10$ units and $\lambda=1.0 $ units.}
\label{fig3}
\end{figure}

No change in the x and y coordinates has also been noted as we change the direction of object beam, keeping the direction of reference beam fixed. However, if we change the direction of reading wave ($cr(1)=0.1;~0.2;~0.3$), keeping the direction cosines of the reference beam fixed ($cr(1)=0.2$), the peak shifts along x-axis. As shown in the Fig. (\ref{fig3}) a shift in peak of about $49\%$ along +ve x-axis and about $54\%$ along -ve x-axis have been observed for first and third cases, respectively. No significant change has been noticed in y-axis.   

\vspace*{1.5cm}

\vspace*{1.5cm}
\begin{figure}[hbt]
\centerline{\epsfig{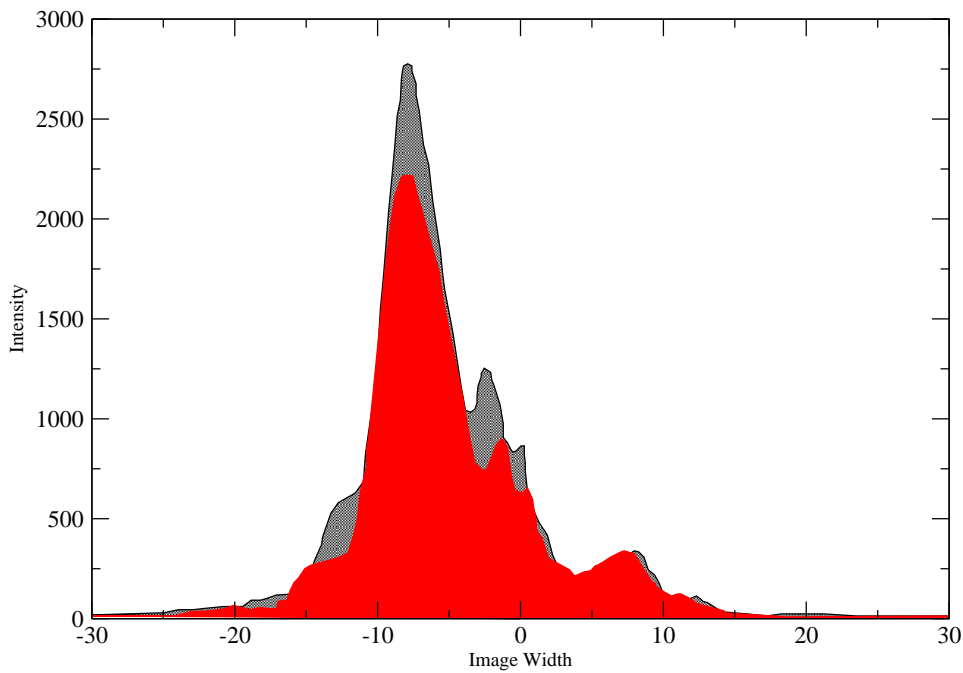}}
\caption{The image for $DD=10$ units, $D=10$ units and $\lambda=0.9;~1.1 $ units.}
\label{fig4}
\end{figure}

In our simulation, there is no significant variation in the off-axis deviation i.e. non-symmetric position of the object.

At last we also investigated the effect of the wavelength on the image formation. As shown in the Fig. (\ref{fig4}), on the variation of $10\%$ in wavelength there is $7-16\%$ up-down shift in the peak.

We have also investigated the sensibility of our holographic mapping on the process 
of changing the wavelength of the reading wave from that of the reference beam, and their direction cosines with a few percent knowing that during experiments this meant destroying the picture. Surprisingly, we did not find this effect in our computations.

\section{Discussion \& Conclusions}
\label{S:conclusions}
The revolution in the Holographic Principle is now a major focus of attention in 
many area of science e.g. gravitational research, quantum field theory and elementary 
particle physics. A popular account of holography can be found in \cite{susskind, susskind_1, jacob, taubes}.

In the present work we discussed a mathematical formalism of creating and reading a hologram, 
in the most general terms. We developed holographic theory for various kind of waves: 
electromagnetic, acoustic, etc. These waves can propagate in several media, solids, 
liquids, plasma, etc. They exhibit a large variety of mathematical structures, such as 
scalar, spinor, vector and tensor. The fields that are neither scalar nor fermions, can 
be Abelian or non Abelian gauge fields if they are massless, but we also considered the 
same fields when non-zero mass is added and so the gauge invariance is in part broken. 

Due to the fact that waves are fields, we borrowed many tools from classical field theory, 
such as Green's function formalism, which is very useful for describing the creation of 
holograms, but it is more important for the understanding of the reading process. This 
formalism can be used for waves of arbitrary spin and waves propagating in a medium. 
As the formalism applies in any dimensions, we presented some conclusions for hyperspace 
and for the two-dimensional space, which is the only case of infra-dimensions that makes sense. 

We first discussed the massive and massless scalar case, then we justified why this 
could be generalized for the vector fields, such as the electromagnetic field in the paraxial 
approximation, and gave a generalized procedure that applies even in the non-paraxial case 
whether the wave is transversal or longitudinal. We also extended this generalized procedure for spinors and tensors.

In our study we obtained the image of a point-like object, like any optical procedure \cite{BW}, a blurry picture instead of a sharp point. Let us bring to your notice that in our previous work \cite{bc} we have conjectured that the uncertainty in the quantum mechanics is due to the holographic basis of physical reality. We 
have argued that for the macro particles (classical objects) this fuzziness, noise and wave 
pattern due to holographic projection are weak and so hard to observe in daily life.  

We investigated the aberrations that are due to a reading beam hitting the screen in a slightly different 
direction that of the reference wave and a slightly different wavelengths, etc. We measured the 
width of image so formed and its dependence on the size of the holographic picture, 
the distance between the screen and the object, the distance of the object from the 
symmetry axis of the screen, the nature of the mapping and reading wave.

In this work we presented the simplest case of scalar wave and a point-like object in 
a two-dimensional space. The infra-space hologram is one dimensional and is read in a two-dimensional space.

In the future work we plan to investigate scalar field in $3d$ and $4d$ space and the electromagnetic fields with  an analogy of the massive case and propagations in a medium, especially in dielectrics and plasmas.
We also plan to describe some possible integrations of the path-integral \cite{path_integral} formalism into our description.

\section{Acknowledgements}
BCC thanks Inter-University Centre for Astronomy \& Astrophysics (IUCAA) Pune for the hospitality provided during the preparation of this work.

\end{document}